\documentclass[twocolumn]{aastex631}

\usepackage{xcolor}
\usepackage{amsmath}

\begin{document}

\title{Microphysics of Particle Reflection in Weibel-Mediated Shocks}

\author{Jasmine Parsons}
\affiliation{Department of Astrophysical Sciences, Princeton University, 4 Ivy Lane, Princeton, NJ 08540}

\author{Anatoly Spitkovsky} 
\affiliation{Department of Astrophysical Sciences, Princeton University, 4 Ivy Lane, Princeton, NJ 08540}

\author{Arno Vanthieghem}
\affiliation{Department of Astrophysical Sciences, Princeton University, 4 Ivy Lane, Princeton, NJ 08540}
\affiliation{International Research Collaboration Center, National Institutes of Natural Sciences, Tokyo 105-0001, Japan}

\begin{abstract}
Particle-in-cell (PIC) simulations have shown that relativistic collisionless shocks mediated by the Weibel instability accelerate $\sim$1\% of incoming particles, while the majority are transmitted through the shock and become thermalized. The microphysical processes that determine whether an incoming particle will be transmitted or reflected are poorly understood. We study the microphysics of particle reflection in Weibel-mediated shocks by tracking a shell of test particles in a PIC simulation of a shock in pair plasma. We find that electrons in positron-dominated filaments and positrons in electron-dominated filaments efficiently reflect off of strong magnetic structures at the shock. These reflected particles headed towards the upstream must then find filaments of the same sign of current as the current carried by the reflected particles in order to successfully move with the shock and participate in diffusive shock acceleration (DSA). The final injection efficiency on the order of $\sim$1\% thus results from the effectiveness of the initial reflection at the shock and the reflected particles’ probability of survival in the upstream post-reflection. We develop a model that predicts the fraction of high-energy particles as a function of the properties of Weibel filamentation.
\end{abstract}

\section{Introduction} \label{sec:intro}

Collisionless shocks are widely believed to be the dominant production mechanism of nonthermal particles in the universe. These shocks are mediated by collective electromagnetic interactions, rather than collisions between particles, and are thus commonly found in astrophysical contexts. Relativistic collsionless shocks occur, for example, in active galactic nuclei (AGN) jets, pulsar wind nebulae (PWNe), and gamma-ray burts (GRBs) \citep[see, e.g.,][]{Koyama_1995, de_Jager_1992, Medvedev_1999, Waxman_2006}. In the case of GRB afterglows, however, the ambient plasma is effectively  unmagnetized, and thus the magnetic fields required for collisionless shock formation must be self-generated.

In unmagnetized shocks, collective interactions are mediated by the Weibel instability \citep{Weibel_1959, Moiseev_1963, Medvedev_1999, Lyubarsky_2006, Achterberg_2007_I, Achterberg_2007_II, Pelletier_2019}. This instability arises from an effective temperature anisotropy in counterstreaming plasmas, and manifests itself as the generation of a series of density filaments along the shock normal containing alternating currents \citep{Bret_2008, Bret_2010a, Lemoine_2011, Shaisultanov_2012}. These currents are associated with kinetic-scale electromagnetic field structures that shape slowdown, heating, and acceleration of particles \citep{Lemoine_2019_II, Lemoine_2019_III}.

Particle-in-cell (PIC) simulations self-consistently capture the highly non-linear interplay between shock structure, electromagnetic turbulence, and particle acceleration.  The PIC method has thus allowed for extensive studies of the properties of relativistic, unmagnetized pair plasma shocks \citep[see, e.g.,][]{kato_2007, Spitkovsky_2008b, Kato_2008, Nishikawa_2009, Martins_2009, Sironi_2011, Haugbolle_2011, Lemoine_2019_PRL, Rajawat_2021}. 
Furthermore, Weibel-mediated shocks have recently begun to be studied in laser-driven experiments, demonstrating experimentally the formation of nonrelativistic shocks and subsequent particle acceleration
\citep[see, e.g.,][]{Park_2015,Huntington_2015, Fiuza_2020}. 

Once the shock has formed, a high-energy tail in the downstream particle energy distribution develops. The dominant acceleration mechanism for incoming particles is thought to be diffusive shock acceleration (DSA), wherein particles gain energy by crossing back and forth across the shock, scattering off of converging magnetic turbulence \citep{Axford_1977, Krimskii_1977, Bell_1978,Blandford_Ostriker_1978}. However, the microphysical details of the first interaction of incoming particles with the shock, i.e., before particles begin participating in DSA, remain unclear. For example, PIC simulations have shown that relativistic Weibel-mediated shocks efficiently accelerate $\sim$1\% of incoming particles via DSA \citep{Spitkovsky_2008b,Sironi_2013}. The physics that determines the fraction of particles which end up with high energies is not understood. Studying particle reflection and transmission through the shock is of interest in this context since particles must first be reflected off of the shock front before being accelerated through the DSA process. Determining the precise microphysics behind particle reflection and transmission can thus explain the physics behind the fraction of high-energy particles. In this work, we develop a simplified model that explains the mechanisms through which particles are selected to become part of the high-energy population by investigating the essential properties of particle reflection.

We begin by describing our PIC simulation setup in \ref{simulation_setup}. We then discuss the behavior of a shell of tracked particles, both as a whole and individual orbits, in Section \ref{shell}, which allows us to develop a simple toy model which captures the microphysics of particle reflection in Section \ref{sec:model}. We obtain a physically motivated estimate for the fraction of high-energy particles in Section \ref{sec:estimate}. Finally, we discuss the consequences of our model and summarize our conclusions in Section \ref{sec:disc_conclu}.

\section{Simulation Results} \label{sec:sim}

\subsection{Simulation Setup}\label{simulation_setup}
We performed our simulation using the electromagnetic PIC code TRISTAN-MP \citep{spitkovsky_2005}, with a setup similar to previous PIC simulations of relativistic collisionless shocks \citep[see, e.g.,][]{Spitkovsky_2008b,Sironi_2009,Sironi_2011}. Namely, to set up the shock, an ``upstream" flow of unmagnetized pair plasma traveling in the $-x$ direction is reflected from a conducting wall at $x = 0$. The counterstreaming of the upstream and reflected streams forms a shock, which propagates away from the wall in the $+x$ direction. We perform our simulation in the downstream rest frame. The simulation domain is a rectangular box in the $xy$ plane, with periodic boundary conditions in $y$. The $+x$ boundary of the box is continuously expanding to save computational resources. Our box has a width of $48\, c/\omega_{p}$, where $\omega_{p} = \sqrt{4\pi e^2 n/\gamma_0 m_e}$ is the upstream relativistic plasma frequency. In this expression, $n = n_{e^+}+n_{e^-}$ is the apparent total pair density, $e$ and $m_e$ are the electron charge and mass, and $\gamma_0$ is the bulk Lorentz factor of the incoming flow in the downstream frame. In our simulation, we use $\gamma_0 = 5$, 32 particles per cell per species, and a spatial resolution of 10 cells per skin depth, $c/\omega_{p}$. Our resulting shock structure and particle spectrum is similar to previous simulations of relativistic, unmagnetized pair shocks \citep[see, e.g.,][]{Spitkovsky_2008b,Sironi_2013}. The simulation is run until time $\omega_pt=1764$, which is long enough to form a power-law tail from DSA, but before the nonlinear evolution of the shock manifests itself \citep[e.g.,][]{Keshet_2009}.

\subsection{Shell of Tracked Particles} \label{shell}

\begin{figure}
  \includegraphics[width=0.5\textwidth]{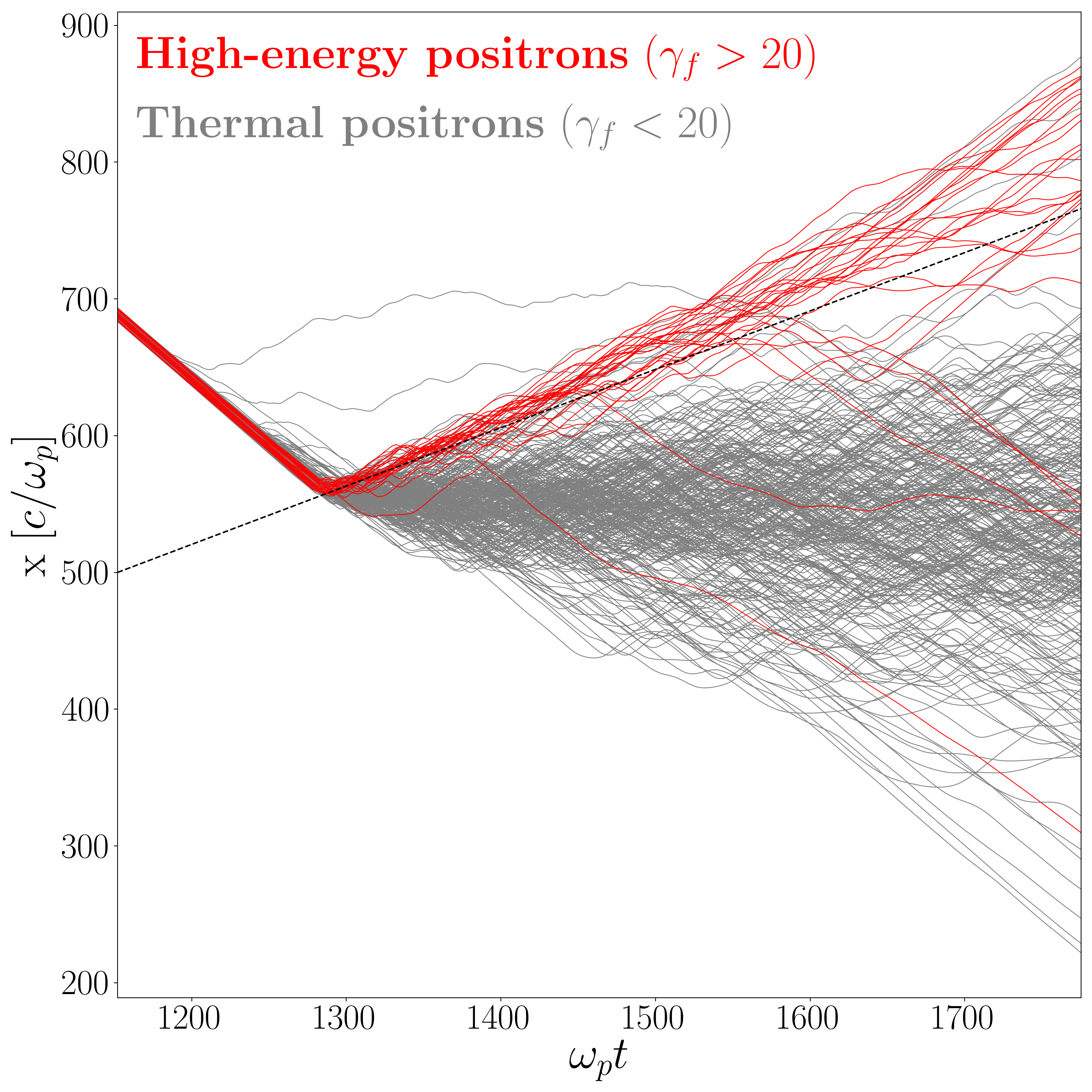}
  \caption{Spacetime orbits of a sample of high-energy positrons (i.e., with $\gamma_f > 20$) shown in red and thermal positrons (i.e., with $\gamma_f < 20$) shown in grey. The approximate location of the shock through time is represented by a black dashed line.}
  \label{ref_pos_spacetime}
\end{figure}

\begin{figure*}
  \centering 
  \includegraphics[width=0.99\textwidth]{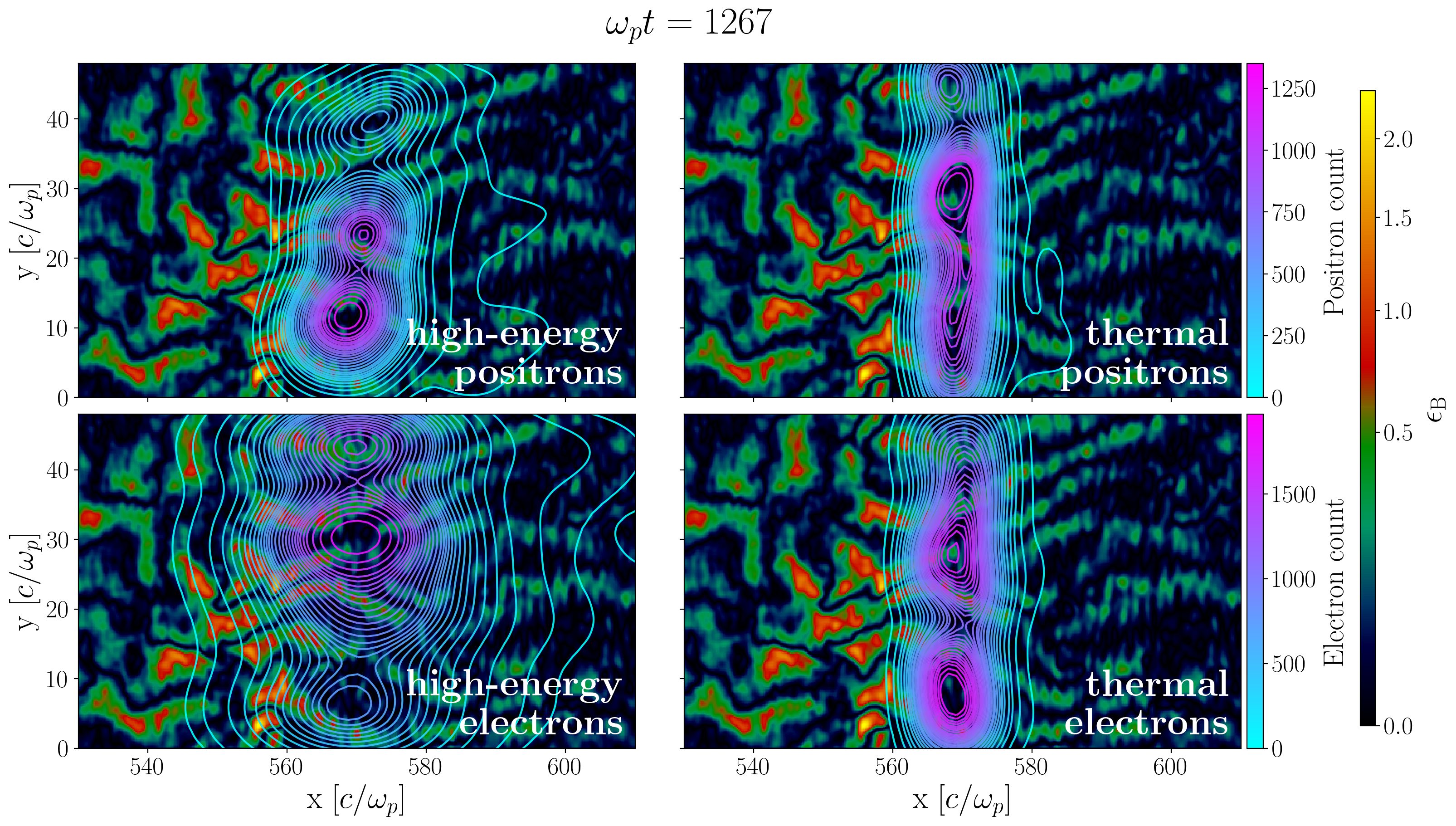}
  \caption{Contours of the four subshells of tracked particles described in Section \ref{shell}, shown at $\omega_pt = 1267$, right before the particles hit the shock. Clockwise from the top left, the contours show high-energy positrons, thermal positrons, thermal electrons, and high-energy electrons. The contours are plotted on top of the magnetic energy fraction $\epsilon_\mathrm{B} = B^2/8\pi \gamma_0 nm_ec^2$. Note that the colorbar for $\epsilon_\mathrm{B}$ is nonlinear. In order to compare the contours of the high-energy particles (left column) and the thermal particles (right column), the total population of thermal particles is downsampled to match the number of high-energy particles of each respective species. In this snapshot, the particles are moving in the $-x$ direction, towards the shock which is located at $x \sim 560\,c/\omega_p$.}
  \label{contour_before_shock}
\end{figure*}

\begin{figure*}
  \centering 
  \includegraphics[width=0.99\textwidth]{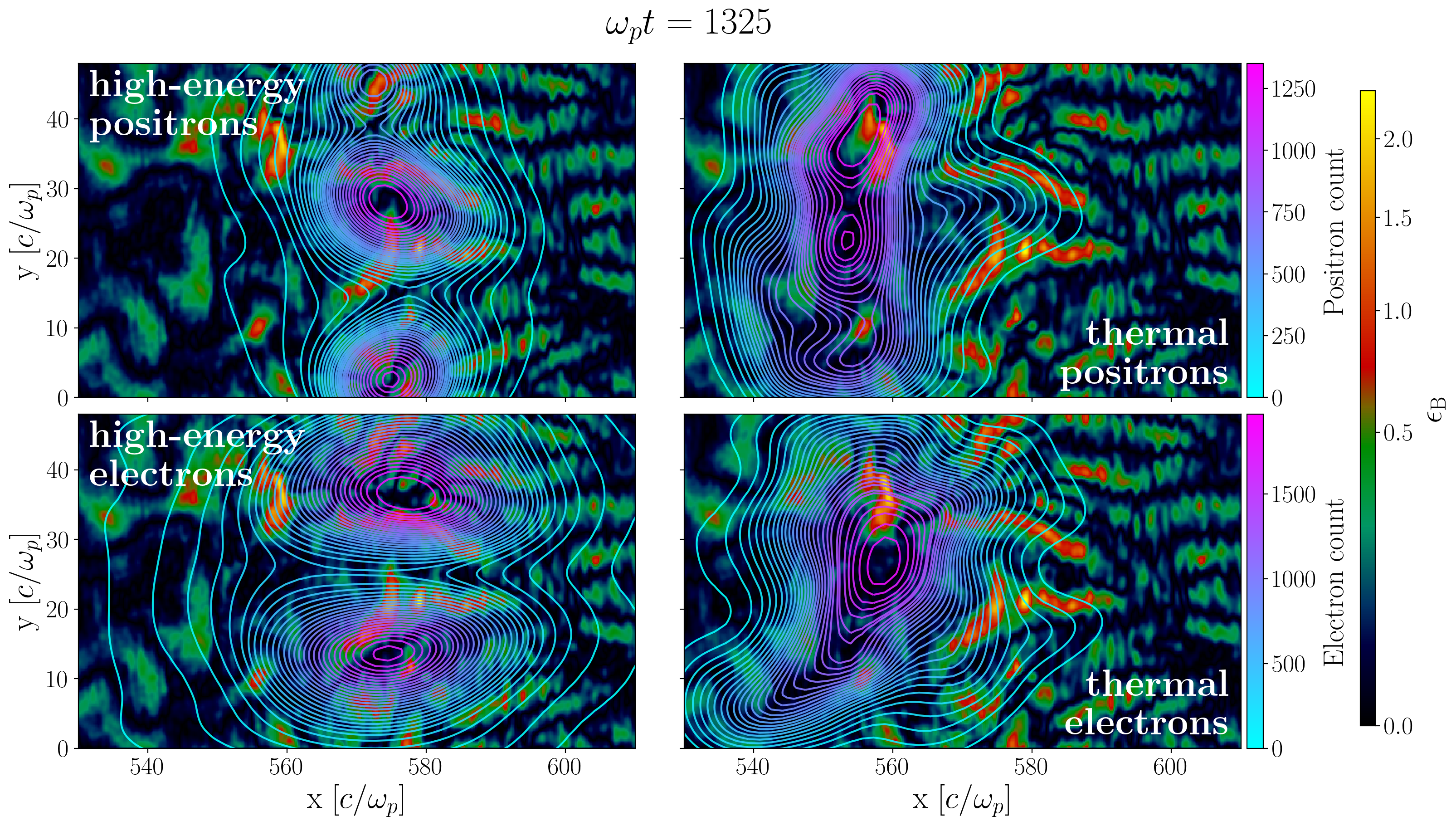}
  \caption{Contours of the four subshells of tracked particles described in Section \ref{shell}, shown at $\omega_pt = 1325$, right after the particles hit the shock. Clockwise from the top left, the contours show high-energy positrons, thermal positrons, thermal electrons, and high-energy electrons. The contours are plotted on top of the magnetic energy fraction $\epsilon_\mathrm{B} = B_z^2/8\pi \gamma_0 nm_ec^2$. Note that the colorbar for $\epsilon_\mathrm{B}$ is nonlinear. In order to compare the contours of the high-energy particles (left column) and the thermal particles (right column), the total population of thermal particles is downsampled to match the number of high-energy particles of each respective species. In this snapshot, the thermal particles (i.e., the right-hand column) have been transmitted through the shock located at $x \sim 575\,c/\omega_p$ and are isotropizing. The high-energy particles (i.e., the left-hand column) have turned around at the shock and are now moving with the shock in the $+x$ direction.}
  \label{contour_after_shock}
\end{figure*}

\begin{figure*}
  \centering 
  \includegraphics[width=0.99\textwidth]{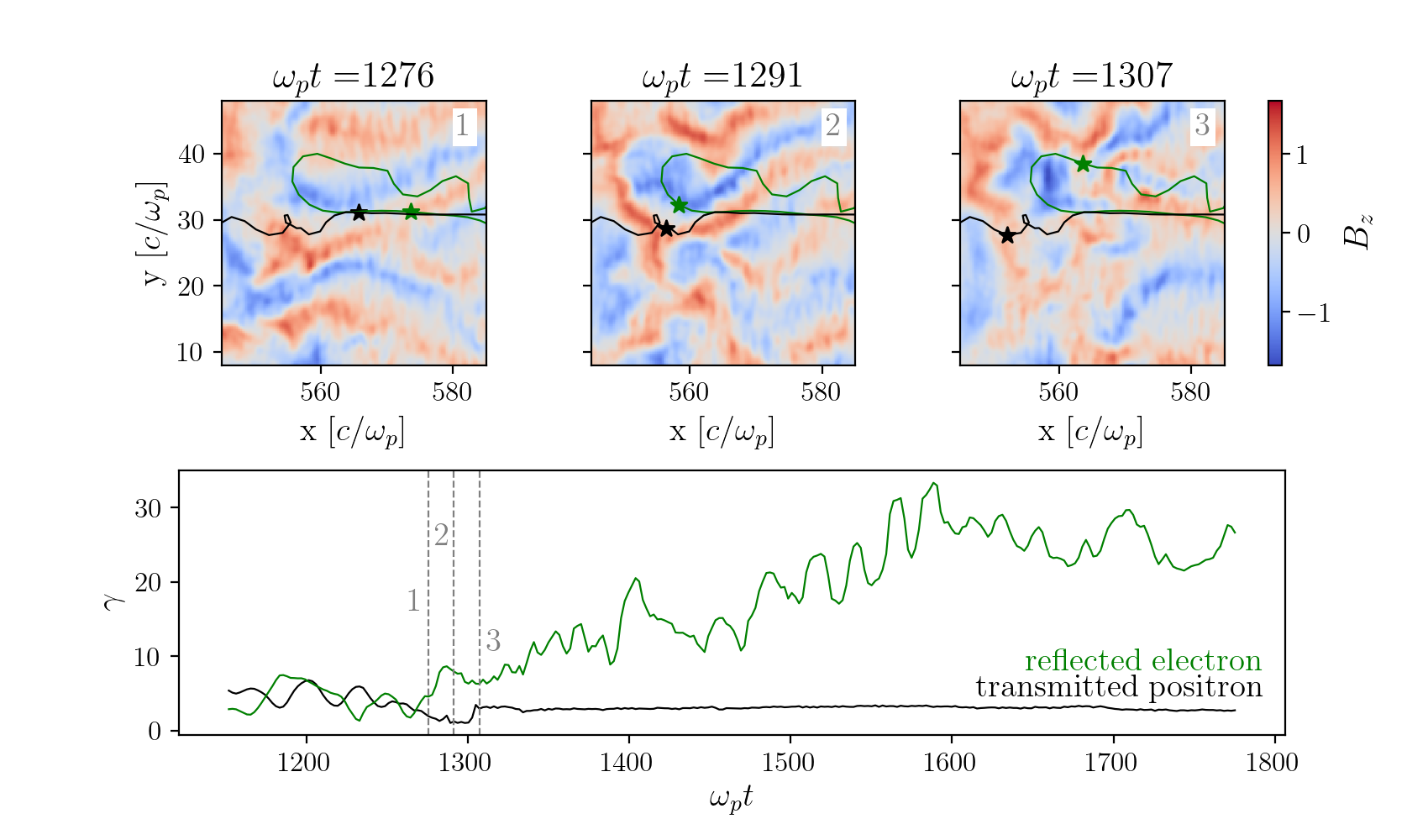}
  \caption{Top row: Snapshots of representative orbits of a positron (black) which is transmitted through the shock and an electron (green) which is reflected by the shock and ends up with high energy. These two particles approach the shock in the same negative current (i.e., positron-dominated) filament. This filament contains both the majority of tracked thermal positrons and high-energy electrons hitting the shock, as seen in Figure \ref{contour_before_shock}. The locations of the particles along their respective orbits at times in the subpanels are shown as stars. The orbits are overplotted on the magnetic field $B_z$ at the time of each snapshot, where $B_z$ is expressed in units of $\sqrt{8\pi \gamma_0 nm_ec^2}$. Bottom row: Energy evolution of the thermal positron (black) and the reflected electron (green). The three grey dashed vertical lines denote the three times corresponding to the snapshots in the top row.}
  \label{orbit_plot}
\end{figure*}

We track a shell of $\sim$160 000 particles of each species. The shell is selected between $162$ $c/\omega_{p}$ and $172$ $c/\omega_{p}$ in front of the shock at $\omega_{p} t = 1152$, to ensure that the shock has had enough time to fully form. We choose a narrow thickness for the shell of $10$ $c/\omega_{p}$ to ensure that particles in the shell hit the shock at the same time. We also ensure that the tracked particles are heading toward the shock for the first time. In other words, our shell does not include any particles previously reflected at the shock. The particles in the shell are randomly selected along the $y$ direction. The shell moves in the $-x$ direction before hitting the shock at around $\omega_{p}t \sim 1276$. We then sort the tracked particles into the categories of `high-energy' and `thermal'. We determine a particle to be `high-energy' if, at the end of the simulation, its energy is high enough for it to be considered part of the nonthermal tail of the downstream particle distribution. We denote the Lorentz factor of a particle at the end of the simulation as $\gamma_f$. After observing the downstream spectrum of all particles at the end of the simulation, we consider that tracked particles with $\gamma_f >  4\gamma_0$ (i.e., $\gamma_f >  20$) are `high-energy', while all other tracked particles are `thermal'. By this definition, we find that $\sim$1.2\% of electrons and $\sim$0.9\% of positrons in our tracked shell are high-energy (the discrepancy is attributed to sample variance). It is also worth noting that we are selecting for all tracked particles with $\gamma_f > 4\gamma_0$, not only those in the downstream, despite using the downstream spectrum to determine this nonthermal cutoff in energy at $\gamma \sim 20$. We use the categories of `high-energy' and `thermal' only as tools to observe differences in behavior between groups of particles. 

It is worth emphasizing that `high-energy' and `reflected' are not synonymous. Indeed, we will later see that a significant fraction of `thermal' particles are reflected off of the shock without ending up in the high-energy tail of the particle distribution. The categorization of tracked particles into `high-energy' and `thermal' rather than `reflected' and `transmitted' is based on the assumption, which we later confirm, that reflection is necessary but not sufficient for joining the high-energy tail of the particle distribution.

To show the basic difference between the high-energy and thermal particles from the tracked shell, in Figure \ref{ref_pos_spacetime}, we plot the spacetime orbits of a sample of high-energy positrons (red) and thermal positrons (grey) from the tracked shell. The approximate position of the shock through time is obtained by following the peak in the transversely averaged magnetic field strength, and is plotted as a dashed black line. Both the future high-energy and thermal positrons travel in the $-x$ direction before hitting the shock at $\omega_pt \sim 1276$, at which point the future high-energy positrons begin moving in the $+x$ direction along with the shock. The thermal particles, which constitute the vast majority of the total particles, are transmitted through the shock and isotropize, coming to a rest in the downstream.

The tracked shell can thus be divided into two subshells for each species, one containing only future high-energy particles and one containing only future thermal particles. We can then observe the behavior of these subshells as a whole throughout the simulation, as well as the behavior of individual particles within these subshells. We begin by comparing these four subshells before and after interacting with the shock. Contour plots of the four subshells on top of magnetic field strength before and after hitting the shock are shown in Figure \ref{contour_before_shock} and Figure \ref{contour_after_shock} respectively.

First, before hitting the shock, particles that will become high energy are not evenly distributed amongst the thermal particles, as can be seen in Figure \ref{contour_before_shock}. Rather, high-energy particles are over-represented in certain local regions, and under-represented in others. These regions of over-representation can be interpreted as local zones of high reflectivity on the shock, and are different for both species. Importantly, the locations of regions of high reflectivity for one species are correlated with the highest-density regions of thermal particles for the other species. In other words, future high-energy electrons are located in positron filaments, and vice-versa.

After the shell hits the shock, the majority of the particles are simply transmitted through the shock to the downstream and become thermalized, as can be seen in the thermal particle panels (right column) of Figure \ref{contour_after_shock}. The future high-energy particles, on the other hand, get turned around by the shock and begin heading in the $+x$ direction, towards the upstream. As can be seen in the high-energy particle panels (left column) of Figure \ref{contour_after_shock}, the high-energy particles are localized in distinct regions between areas of strong magnetic field, with the high-energy electrons and positrons being anti-correlated. These regions correspond to the locations of incoming upstream filaments with currents matching those produced by the high-energy particles as they move towards the upstream. Being in these regions of current allows the high-energy particles to continue moving in the $+x$ direction along with the shock, rather than being advected into the downstream, as was pointed out in \cite{kato_2007}. In other words, in order to move with the shock, electrons must find a positron-dominated filament crashing into the shock (and vice versa for positrons). By moving with the shock, the particles are then able to gain energy. However, since the upstream filaments merge and kink as they hit the shock, the future high-energy particles must split up and repeatedly find filaments until they are sufficiently accelerated to move parallel to the shock front or escape further into the upstream.

Indeed, in every successive change of filament, many particles are unable to find the next filament of the right current sign and are advected into the downstream without sufficient energy to be considered part of the high-energy tail of the distribution. This can be seen in Figure \ref{contour_after_shock}, where there are a significant number of `thermal' particles in the same location as the `high-energy' particles: for example, in the top right panel showing contours of the thermal positrons, there are clearly positrons which are trapped in the positive current region at $y \sim 27\,c/\omega_p$ and $x \sim 575\,c/\omega_p$ and are moving to the right. These positrons were reflected when they hit the shock and found an upstream electron-dominated filament which allowed them to carry along with the shock for a while. However, when this electron-dominated filament split, most of these positrons were unable to find another electron-dominated filament and were advected downstream without gaining much energy. This is why they are considered to be in the `thermal' category, as per the definition laid out at the beginning of this subsection. This process of acceleration post-reflection is discussed in further detail in Section \ref{sec:estimate}.

We have now made qualitative observations about the behavior of this shell of particles as a whole before and after hitting the shock. These observations inform our subsequent discussion of the microphysics behind particle reflection and transmission. Indeed, from this shell, we observe that particles are most likely to reflect when they are located in regions of high density of the opposite species when hitting the shock. In order to understand the reason behind this, we can compare the individual orbits of two particles that hit the shock at around the same place and time: a positron that is transmitted through the shock, and an electron that reflects off of the shock. These orbits were chosen to be representative of the behavior of the hundreds of particles whose orbits we observed, and are shown in Figure \ref{orbit_plot}. We focus on the moments right before, during, and right after reflection. These particles are contained within the largest blob of both high-energy electrons and thermal positrons, located at y $\sim 30 \,c/\omega_p$ in Figure \ref{contour_before_shock}.

Both particles approach the shock in a positron-dominated filament, i.e., in a region of negative current. The two particles both hit strong magnetic features at the shock, and have roughly similar energies when hitting the shock, as shown in the bottom panel of Figure \ref{orbit_plot}. However, due to the magnetic field configuration generated by the negative current, the positron is forced to bounce between the strong magnetic features and is unable to turn around. The electron, on the other hand, is able to be flung back towards the upstream. Once reflected, the electron begins gaining energy, before ending the simulation with an energy of $\gamma_f \sim 27$, thus entering it into the `high-energy' category. We note that while the specific reflected electron shown in Figure \ref{orbit_plot} hits the shock with slightly more energy than the transmitted positron, there are many examples of reflected particles hitting the shock with even lower energy than their transmitted counterparts, while still being able to end up in the high-energy category at the end of the simulation.

In order to reflect, it is thus insufficient to simply experience certain electromagnetic field strengths above some threshold value, or to hit the shock with a certain energy. Rather, there exist local structures on the shock that are efficient at reflecting one species of particles, and conversely efficient at funneling the other species of particles into the downstream. In the next section, we develop a simplified toy model of these structures, and show that test particles in this simplified electromagnetic setup capture the basic properties of the tracked particle orbits.

\section{Modeling Particle Reflection} \label{sec:model} \label{toy_model_section}

\begin{figure}
  \includegraphics[width=0.49\textwidth]{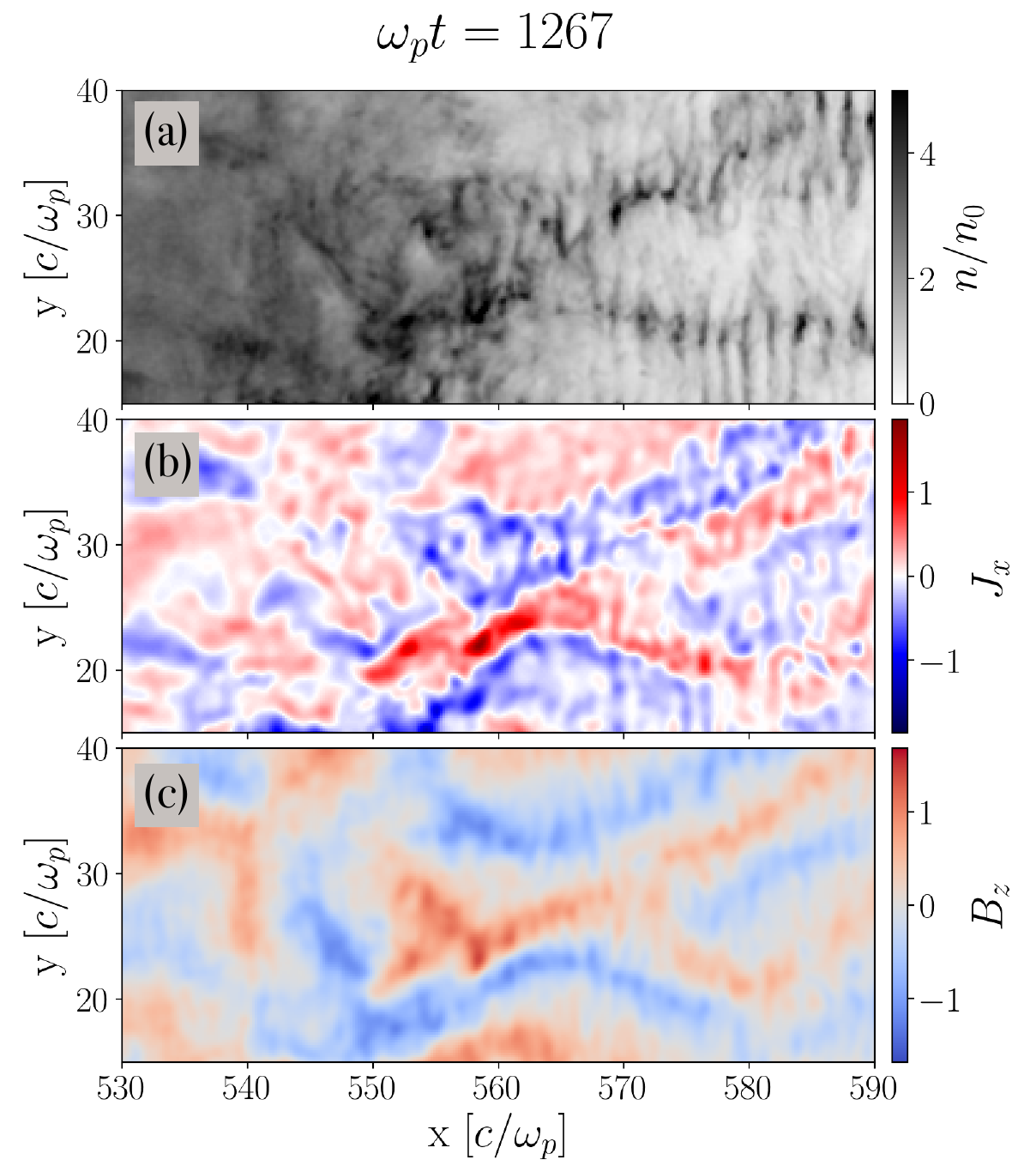}
  \caption{Snapshot of (a) normalized density, (b) current $J_x$ expressed in units of $\omega_p \sqrt{8\pi \gamma_0 nm_ec^2}/4\pi$, and (c) magnetic field $B_z$ expressed in units of $\sqrt{8\pi \gamma_0 nm_ec^2}$, taken at $\omega_pt = 1267$. The top panel shows two distinct density filaments approaching each other at the shock, which is located at $x \sim 560\,c/\omega_p$. The middle panel shows the negative and positive regions of current created by the two density filaments. The magnetic field generated by these current regions is shown in the bottom panel.}
  \label{dens_Jx_Bz_figure}
\end{figure}

\begin{figure}
  \includegraphics[width=0.5\textwidth]{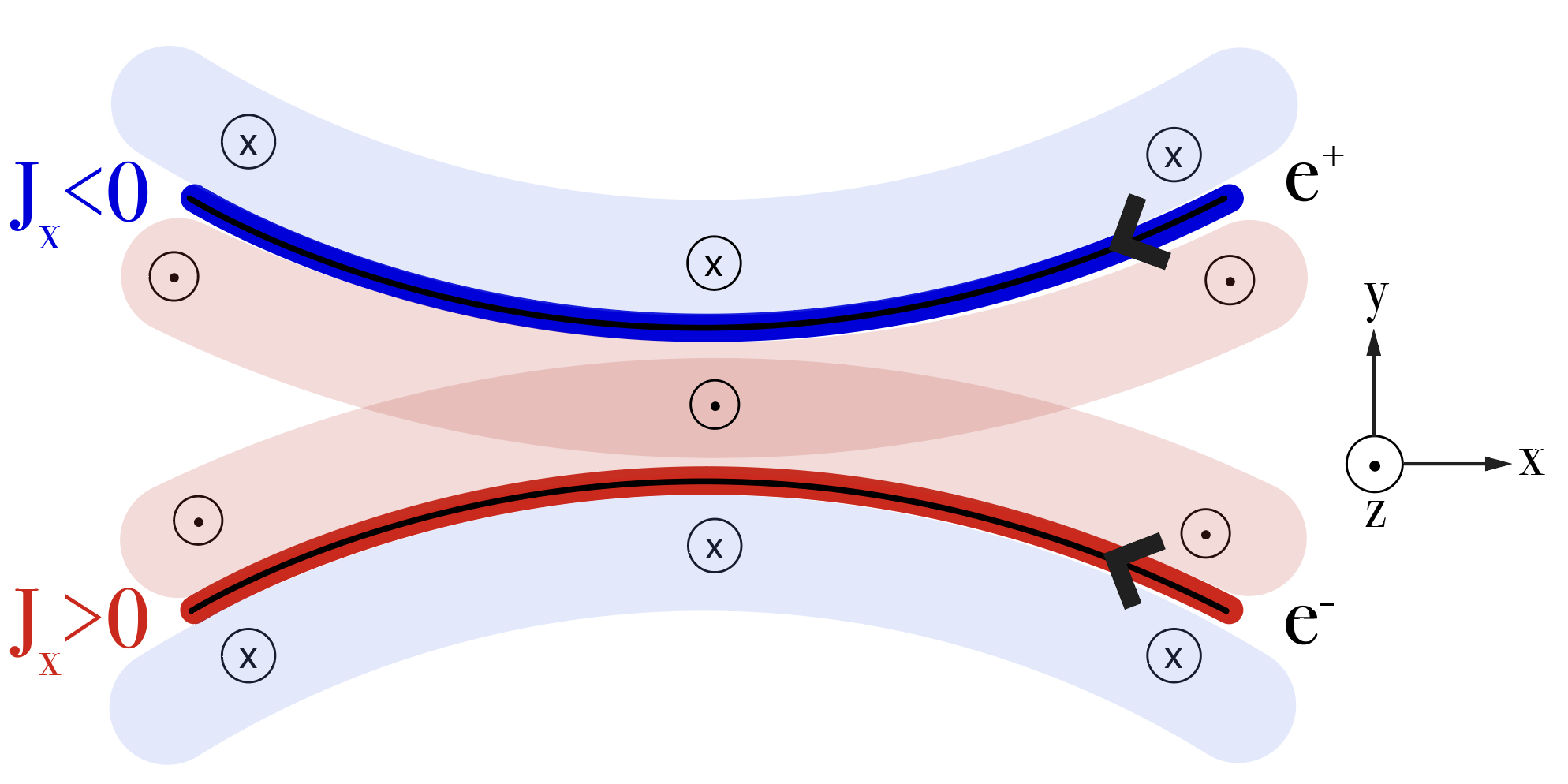}
  \caption{Sketch of toy model described in Section \ref{toy_model_section}. The black curves represent density filaments dominated by positrons (top) and electrons (bottom). The dark blue and dark red curves respectively represent regions of negative and positive current around the density filaments. These current regions generate magnetic fields around them. Positive (out of the page) and negative (into the page) magnetic regions are shown as light red and light blue shaded regions, respectively.}
  \label{toy_model_figure}
\end{figure}

We begin by discussing a toy model which captures the microphysics of particle transmission and reflection. Namely, we investigate the behavior of particles in a particular magnetic field configuration created by the approach of a positron filament and an electron filament. While in the upstream the filaments that merge are mainly of the same sign of current and dominated by the same charge species, the nonlinear filament evolution near the shock can force filaments of opposite currents and species to come close together. The resulting magnetic field structure is commonly seen at the shock in our simulation, as shown in Figure \ref{dens_Jx_Bz_figure}. This figure shows a snapshot of density (top panel), current (middle panel), and magnetic field (bottom panel) at $\omega_pt = 1267$, the same time as Figure \ref{contour_before_shock}, right before the tracked particles hit the shock at $x \sim 560\,c/\omega_p$. This structure can be identified in Figure \ref{contour_before_shock} as a highly reflective region on the shock, since the dominant clump of high-energy electrons (at $y \sim 30\,c/\omega_p$) and a significant clump of high-energy positrons (at $y \sim 22\,c/\omega_p$) hit the shock at that location. 

We use Figure \ref{toy_model_figure} to illustrate the typical structures observed in the simulation (see Figure \ref{dens_Jx_Bz_figure}), showcasing schematic representations of the density, current, and magnetic field profiles. As the electron filament with positive current (bottom curve, in dark red) and the positron filament with negative current (top curve, in dark blue) are forced to approach each other at the shock, the magnetic field around them is amplified. 
In this picture, positrons heading towards the downstream (i.e., in the $-x$ direction) within the region of negative current would bounce between the positive $B_z$ region (middle red-shaded region) and the negative $B_z$ region (top blue-shaded region). These positrons approaching the shock in a positron filament are then advected downstream, unable to reflect and participate in DSA. 
Electrons in this negative current region, on the other hand, would either be flung in the $-y$ direction if they hit the positive $B_z$ region (middle red-shaded region) or in the $+y$ direction if they hit the negative $B_z$ region (top blue-shaded region). These electrons approaching the shock in a positron filament would thus have the opportunity to fully turn around at the shock and start heading back in the $+x$ direction, towards the upstream. The same situation is true in reverse for the electron filament with positive current.

\begin{figure}
  \includegraphics[width=0.5\textwidth]{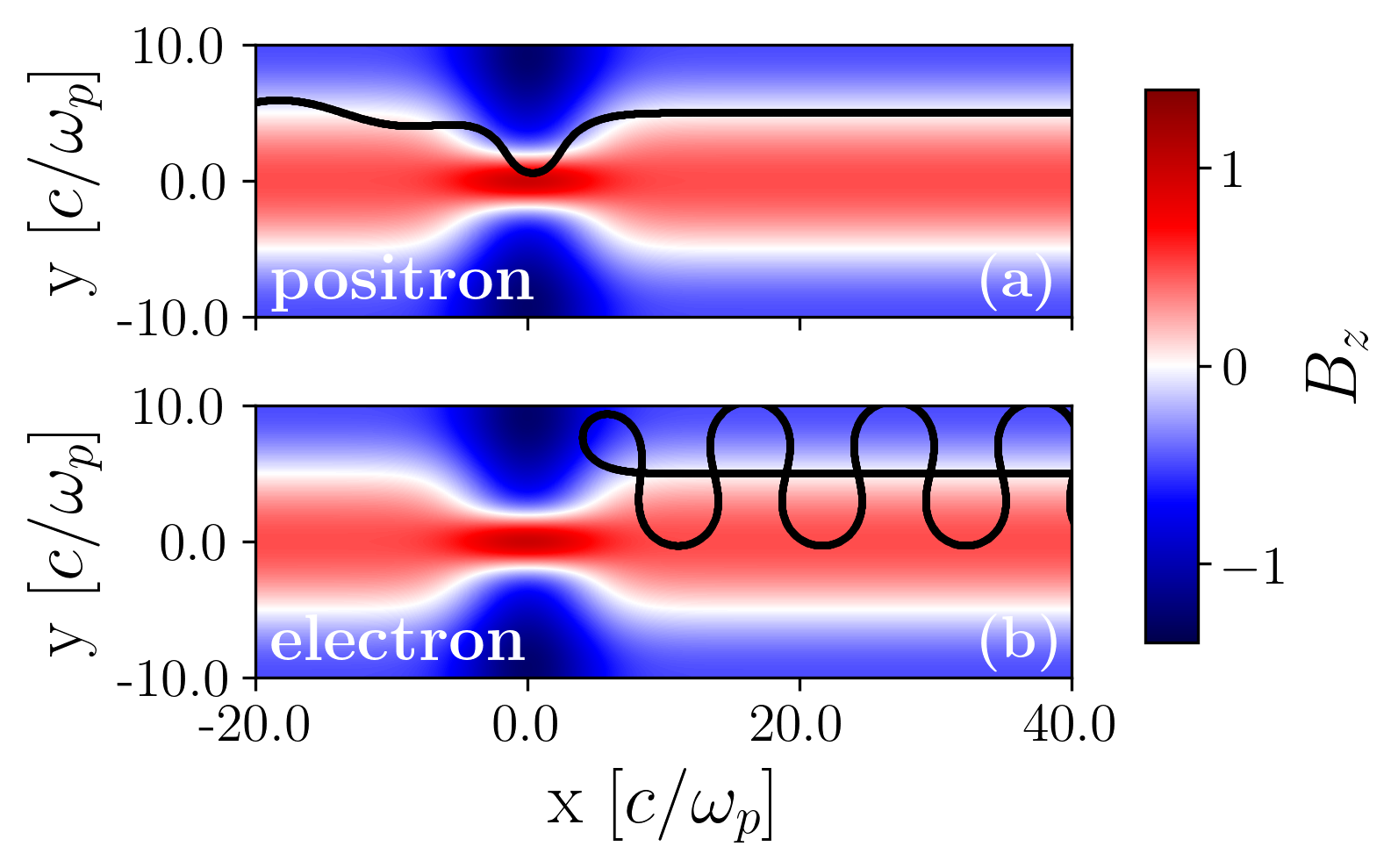}
  \caption{Orbits of two representative test particles, one positron (panel (a)) and one electron (panel (b)), in the electromagnetic field of the toy model described in Section \ref{toy_model_section}. The orbits are plotted on top of $B_z$ where the normalized magnetic field is sign($B_z$)$\sqrt{\epsilon_\mathrm{B}}$, and no electric fields are included. Both particles begin in a positron-dominated filament traveling in a straight line along the $-x$ direction and approach the region of $B_z$ amplification at $x \sim 0\,c/\omega_p$. The positron is transmitted through the region while the electron is reflected off of the amplification region and returns upstream.}
  \label{int_orbit_plot}
\end{figure}

We illustrate this qualitative behavior in the test-particle limit in Figure \ref{int_orbit_plot}.  We impose a static sinusoidal magnetic field perturbation with a wave vector along the $y$-axis and impose a kink of the current filament at $x = 0\,c/\omega_p$.
In Figure \ref{int_orbit_plot}, we sample orbits of a positron (top panel) and an electron (bottom panel) which start out in a positron filament located at $y = 5\,c/\omega_p$. When the positron hits the region of $B_z$ amplification and compression at $x \sim 0\,c/\omega_p$, the particle bounces in the filament and is transmitted through the region. The electron, on the other hand, moves straight before reflecting off of the region of $B_z$ amplification, and then travels upstream within the positron filament with negative current. This same physical picture is true in reverse for the electron filament located at $y = -5\,c/\omega_p$. Even in this very simplified field setup, the basic properties of the orbits of the transmitted positron and reflected electron shown in Figure \ref{orbit_plot} are recovered. Of course, in the full simulation, the incoming density filaments are changing dynamically. Thus, the reflected particle would not necessarily be trapped in the same filament that brought it to the shock, as is shown in Figure \ref{int_orbit_plot}, but instead might have to find another filament of the same sign of current along the shock.

It is worth noting that both positrons and electrons in the negative current region could experience the same magnitude of $B_z$. Indeed, when analyzing particle orbits, we found no consistent difference in the magnitude of electromagnetic fields felt by high-energy and thermal particles when hitting the shock. It is instead the location of the particle within the magnetic structure shown in Figure \ref{toy_model_figure} that determines whether the particle will be reflected by the shock or transmitted. 

In this reduced description, particles approaching the shock in filaments dominated by particles of the opposite sign are reflected off of the shock. Particles in filaments dominated by particles of the same sign, on the other hand, are transmitted through the shock.
The locations of the clumps of particles of each species which become high energy, shown in the left column of Figure \ref{contour_before_shock}, now follow naturally from the toy model in Figure \ref{toy_model_figure}. Indeed, the clumps of high-energy electrons approach the shock in regions of negative current, while the high-energy positrons approach the shock in regions of positive current. This toy model thus explains the correlation between the locations of reflection for one species and the highest-density regions of thermal particles for the other species, as seen in Figure \ref{contour_before_shock} and discussed in subsection \ref{shell}.

In addition, to understand why certain clumps contain more high-energy particles than others, we note that not all current filaments have the same density. Indeed, there are regions of nonzero current in between density filaments. For example, in Figure \ref{dens_Jx_Bz_figure}, the positron filament (top filament) and electron filament (bottom filament) shown in the top density panel have corresponding negative (blue) and positive (red) regions of current, shown in the middle panel. There are regions of current shown in the middle panel, however, that are not associated with a density filament in the top panel, such as the positive current region at $y \sim 38\,c/\omega_p$ and the negative current region at $y \sim 20\,c/\omega_p$. These regions of nonzero current between density filaments can still show similar magnetic structure properties to those shown in Figure \ref{toy_model_figure}, and are still reflective. In other words, any two approaching opposite current regions lead to a reflective structure. However, some structures are created by currents associated with density filaments, and will thus simply be hit with more incoming particles, leading to a larger number of reflected particles.

We now discuss and estimate the parameters of our reduced description for particle reflection. The different stages of particle reflection rely, first, on the fraction of incoming particles undergoing a strong reflection at the shock, and then, on the properties of the subsequent scattering events across the shock transition. 

\section{Estimating the high-energy fraction} \label{sec:estimate}

Now that the dominant microphysical characteristics of particle reflection and transmission have been explored, we turn to obtaining an estimate for the percentage of particles which end up in the high-energy tail of the particle distribution. To summarize the physical picture so far, more than $\sim$1\% of incoming particles are reflected off the shock. In order to continue moving in the $+x$ direction along with the shock, reflected particles must find a `channel' of the same sign of current as the particles, i.e., an incoming upstream filament of the opposite species. However, the returning particles cannot simply remain in one channel of the same current forever, since the upstream filaments are constantly kinking and merging at the shock. Once its channel disappears, if the particle does not yet have enough energy to move into the upstream or parallel to the shock surface, it must find another channel in order to move with the shock. Otherwise, the particle is advected downstream, and cannot become part of the $\sim$1\%. In order to arrive at an estimate of the high-energy fraction, we must first estimate the percentage of particles which are initially reflected at the shock. Then, we can estimate the percentage of reflected particles which are able to remain with the shock long enough to become high energy.

\subsection{Initial Reflection}\label{init_reflect_section}

Following observations of the behavior of the tracked shell in Section \ref{sec:sim} and the discussion of the toy model presented in Section \ref{sec:model}, we can now state that, to a good approximation, every incoming positron in an electron filament and every electron in a positron filament reflects when hitting the shock. 
Estimating the percentage of particles of one species in filaments of the opposite species will thus yield the overall percentage of initially reflected particles.

To compute the typical density profile of the particles at the shock transition, we consider a quasistatic equilibrium in an isothermal multi-species system corresponding to electrons and positrons of the incoming and reflected particles~\citep{Vanthieghem_2018}. Hereafter, we refer to the incoming particles as the `background' and the reflected particles as the `beam'. Neglecting the longitudinal modulations of the filaments, the density profile is then obtained assuming a drifting Jüttner-Synge distribution in infinite filaments with out-of-plane magnetic field~\citep{Kocharovsky_2010}:
\begin{align}
    f_s(y, \bm{p}) \,=\, n_{0,s} &\exp\left\{ - \frac{\gamma_s}{T_s} \left[ m_s \gamma(\bm p) + q_s \phi(y)  \right. \right. \nonumber \\
    &\left. \left.- \beta_s \left( p_x + q_s A_x(y) \right)   \right]\right\} \,,
\end{align}
where $(\phi,A_x, 0, 0)$  is 4-vector potential, $T_s$ is the species temperature, $q_s$ is the charge, and $\gamma_s = 1/\sqrt{1- \beta_s^2}$ is the bulk Lorentz factor of the incoming flow. This leads to a proper density profile $n_s$ of the form:
\begin{equation}
    n_s \,=\, n_{0,s} \exp\left\{ - \frac{\gamma_s q_s}{T_s} \left[   \phi(y) - \beta_s  A_x(y)   \right]\right\} \,.
\end{equation}
From Ampère-Maxwell and Gauss-Maxwell equations together with the temperature, velocity, and density of the different components, one obtains the full nonlinear profile of the different species (see~\citealt{Vanthieghem_2018}). The strength of the filament or nonlinearity level reduces to a single parameter
\begin{equation}
    \xi \,\sim\, \left| \frac{\gamma \beta }{T}  \max A_x \right|
\end{equation}
as expressed in the frame where $\phi\to0$. In this frame, we rewrite the density profile as:
\begin{align}
    n_s \,=\, n_{0,s} \exp\left[ - \frac{q_s}{e} \xi \cos\left( 2 \pi y/\lambda \right) \right].
\end{align}
A filament typically extends in the transverse direction over half a period set by the wavelength $\lambda$. Integration over the filament profile leads to the following expression:
\begin{align}
    \int^{3/4}_{1/4}  \,n_s \,{\rm d}\frac{y}{\lambda}\,=\, \frac{n_{0,s}}{2} \, \left[ I_0(\xi) + \frac{q_s}{e} L_0\left( \xi \right) \right]
\end{align}
where $I_0$ is the modified Bessel function of the first kind, and $L_n$ is the modified Struve function, such that the fraction of incoming electrons in a filament dominated by positrons is approximately:
\begin{align}
\label{eq:xi}
    f_{-}\,=\,\frac{n_{-}}{n_{+} + n_{-}} \,=\, \frac{1}{2} \left( 1 - \frac{L_0\left( \xi \right)}{I_0(\xi)} \right)\,\simeq\,\frac{1}{2} - \frac{\xi}{\pi} \,.
\end{align}

The nonlinearity level of the plasma is bounded above by the transition between transverse modes responsible for filament merging and longitudinal kink-unstable modes. In relativistic symmetric pair plasma flow, the transition occurs around $\xi$ of order unity \citep{Vanthieghem_2018}. In unmagnetized relativistic pair shocks, it was shown in \cite{Pelletier_2019} that $\xi$  of the order of a fraction of unity is observed at the shock transition. In what follows, we will assume $\xi\,\sim\,1/2$, with little effect on the final estimates. 

\begin{figure}
  \includegraphics[width=0.5\textwidth]{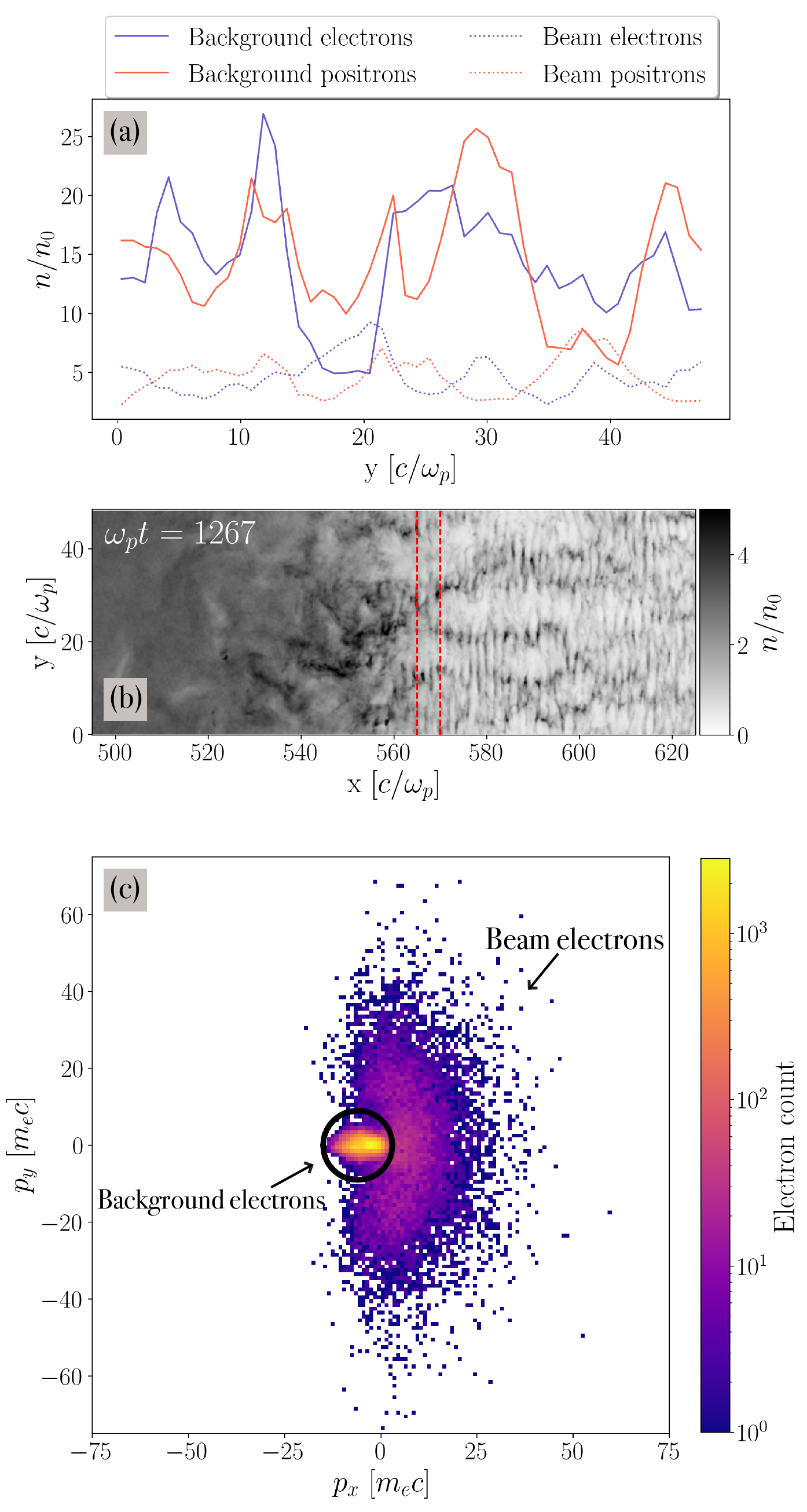}
  \caption{Panel (a): Density profile at $\omega_pt = 1267$ of background electrons (solid purple), background positrons (solid orange), beam electrons (dotted purple), and beam positrons (dotted orange), within a narrow slice in front of the shock. Panel (b): Snapshot of density at $\omega_pt = 1267$, showing the region (delineated by dashed red lines) whose density profile is plotted in panel (a). Panel (c): Density distribution in momentum space of all electrons in the region delineated by dashed red lines in panel (b). Background electrons correspond to the compact yellow population, moving in the $-x$ direction. Beam electrons, on the other hand, correspond to the hot and dilute purple cloud.  Background electrons are thus located within the black circle of radius $9\,m_ec$, while beam electrons are located outside of the black circle.}
  \label{density_profile}
\end{figure}

The expression (\ref{eq:xi}) for the percentage of particles of one species in a filament of the opposite species has been derived assuming a quasistatic equilibrium. We now justify this simplified picture by comparing the density profiles of the background species obtained from the simulation.
In Figure \ref{density_profile}, we plot the density profile of the background and beam particles of both species (panel (a)). Note that this density profile is obtained not from our tracked shell, but rather from all particles in the region delineated by the dashed red lines in panel (b). We separate the background particles from the beam particles by observing the momentum space density distribution for electrons and positrons in the delineated region in front of the shock (panel (c)). We use the diameter of the compact region moving in the $-x$ direction to define a circular region in momentum space. Particles located inside the circular region are determined to be the background particles, and particles located outside the circular region are designated as the beam particles.
We observe that the density profile from the simulation is consistent with a charge fraction of the order of $f_{-}\,\sim\,0.35$ corresponding to $\xi\,\sim\,1/2$. We illustrate the density profile of the background particles in panel (a) of Figure \ref{density_profile}, where the fraction $f_{-}$ corresponds the density of background electrons (solid purple curve) divided by the total density (sum of solid purple curve and solid orange curve) in a positron-dominated filament, seen for example at $y\sim30\,c/\omega_p$. By symmetry, we also take the fraction of incoming positrons in a filament dominated by electrons to be {$f_+ \sim 0.35$}.

In sum, we find that $\sim$35\% of incoming particles are initially reflected at the shock. In the following subsection, we describe our estimation of the percentage of initially reflected particles which are able to become high-energy particles.

\subsection{Subsequent Survival}\label{survival_section}

\begin{figure*}
  \includegraphics[width=0.99\textwidth]{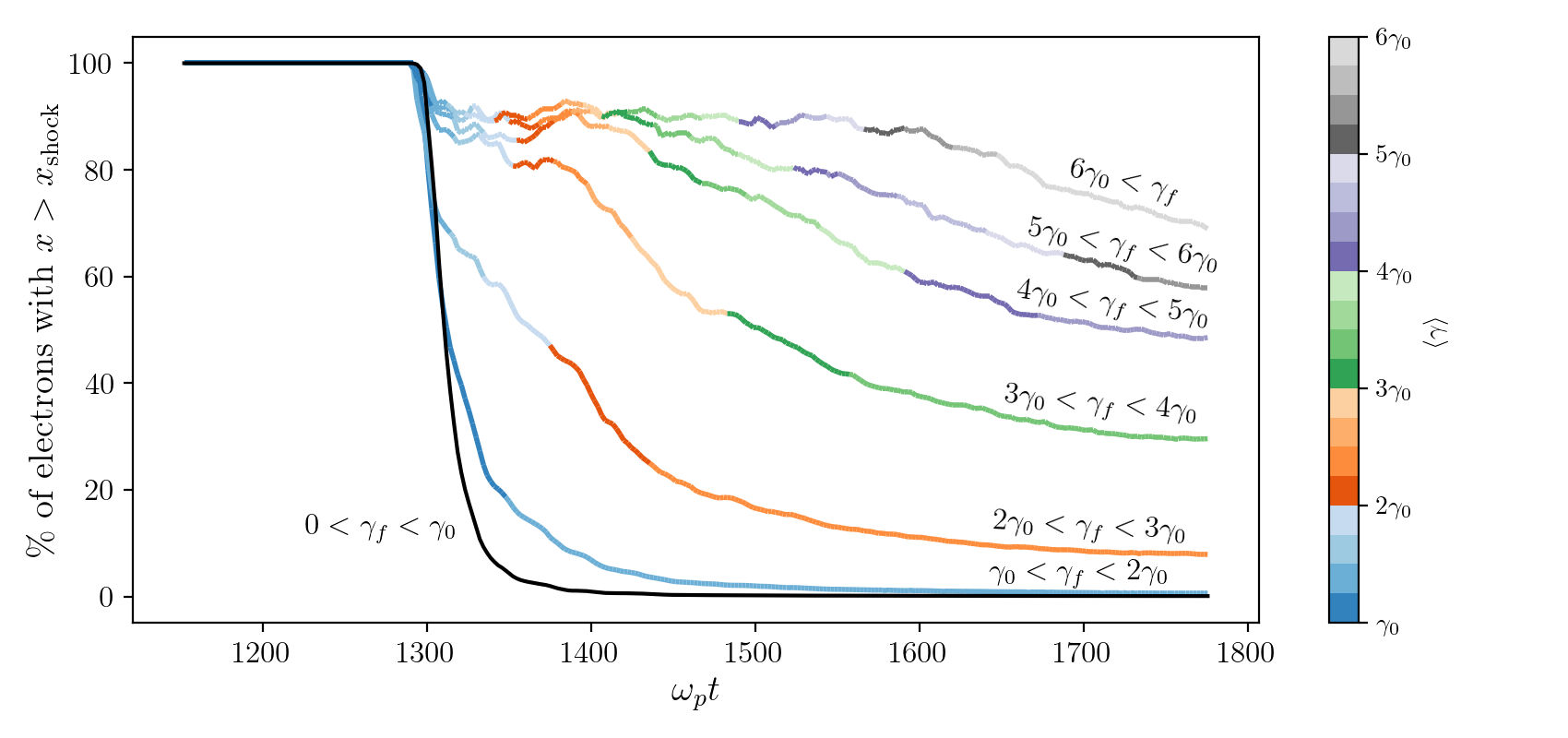}
  \centering 
  \caption{Evolution of the percentage of electrons with $x > x_{\mathrm{shock}}$ through time. The electrons are separated into seven bins based on their final energy $\gamma_f$ in increments of five starting at $\gamma_f = 0$, each of which has its own plotted curve. The curves representing the evolution of the five bins which have $\gamma_f > \gamma_0$ (i.e., all bins except the $0 < \gamma_f < \gamma_0$ bin) are colored by the average energy $\langle \gamma \rangle$ of the electrons in the bin at that time. Note that the color bar only extends to $\langle \gamma \rangle = 6\gamma_0$ for visualization purposes, and that the $6\gamma_0 < \gamma_f$ bin reaches a higher average energy towards the end of the simulation. We note that the equivalent figure for positrons instead of electrons is qualitatively similar.}
  \label{percentage_plot}
\end{figure*}

In order to estimate the percentage of particles in the reflected beam which are able to become high-energy, we separate particle acceleration into two phases. The first, which we call the `pre-acceleration' phase, describes the time during which reflected particles are traveling along with the shock in the $+x$ direction, aided by finding filaments of the same sign of current, and gradually gaining energy. Some particles survive long enough to gain sufficient energy such that they can no longer be bound by the current filaments and thus enter the `free-streaming' phase. During this latter phase, the particles are able to travel farther upstream, as well as parallel to the shock, resulting in significant energy gains through participation in DSA. All particles which constitute the high-energy tail of the particle distribution, i.e., the $\sim$1\%, manage to reach the free-streaming phase. Many particles in the pre-acceleration phase are advected downstream before reaching the free-streaming phase, thus forming the transition between the Maxwellian and the nonthermal tail of the downstream particle distribution.

We have so far made qualitative observations concerning the behavior of the reflected beam informed by the contours of the four subshells of tracked particles shown in Figure \ref{contour_after_shock}. To link our qualitative observations to the two phases described above, we quantify the percentage of particles which move along with the shock through time. In Figure \ref{percentage_plot}, we separate all the tracked electrons into seven bins based of their final energy at the end of the simulation, and plot the time evolution of the fraction of electrons in each energy bin that are located ahead of the shock. Splitting the electrons into separate bins based on their final energy allows us to observe the link between final energy and location with respect to the shock. We emphasize again that the bins are separated by final energy, and not by instantaneous energy, and that the bins contain different numbers of electrons. In addition, the curves associated to bins which have gained energy by interacting with the shock (i.e., $\gamma_f > \gamma_0$) are colored by the instantaneous average energy of the electrons in that bin. This allows us to see when particles in a particular bin gain their energy, and whether particles in different bins gain their energy at the same times.

In Figure \ref{percentage_plot}, at first, all populations begin at 100\%, since the shell is in front of the shock moving in the $-x$ direction. Then, when the shell hits the shock, the electrons which will end with very low energies (solid black line) experience an exponential drop in the percentage of electrons located to the right of the shock, since they pass right through the shock without reflecting.  A larger percentage of the electrons stay with the shock for a longer period of time for each sequential step in final energy, as expected. For electrons with final energies of $\gamma_f < 3\gamma_0$, the drop off in the population moving along with the shock is considerable even at early times. The four curves representing the bins with $\gamma_f > 3\gamma_0$, on the other hand, do not exhibit this drop off at early times, and instead exhibit a similar `plateau'. This plateau indicates that the vast majority (over $\sim$85\%) of electrons which have final energies of $\gamma_f > 3\gamma_0$ move steadily along with the shock for a considerable period of time. In addition, electrons with $\gamma_f > 3\gamma_0$ all exhibit similar, continuous energy gains until the end of the shared plateau at $\omega_pt \sim 1420$. The differences in energy gains after $\omega_pt \sim 1420$ can be attributed to differences in the times particles spend participating in DSA. The existence of this plateau during the time period $\omega_pt \sim 1270 - 1420$ for curves with $\gamma_f > 3\gamma_0$ is thus a signature of the pre-acceleration phase.

The most important takeaway from Figure \ref{percentage_plot} is that the potential of a given particle to gain enough energy to join the $\sim$1\% correlates strongly with its capacity to move with the shock for a significant period of time. Estimating the percentage of particles that are able to move with the shock throughout the entire pre-acceleration phase and reach the free-streaming phase will thus physically explain the $\sim$1\% of particles in the high-energy tail of the particle distribution.

We first calculate the approximate energy at which reflected particles move from the pre-acceleration phase to the free-streaming phase. This occurs when the Larmor radius of the particle becomes comparable to the typical transverse size of the filaments. The relativistic Larmor radius is defined as $r_L \equiv \gamma_0 m c^2 /eB$. Using this formula, with our observed values of the average maximal magnetic field strength in a given filament and a typical filament scale of $\sim$$4$ $c/\omega_p$, we obtain an energy of $\gamma \approx 16 \approx 3.2\gamma_0$. The filament scale is extracted within the shock transition region (seen, for example, at $x \sim 575\,c/\omega_p$ in Figure \ref{contour_after_shock}). This energy threshold of $\gamma \approx 3.2\gamma_0$ is consistent with the signature of the pre-acceleration phase seen in Figure \ref{percentage_plot}, namely that curves with $\gamma_f > 3\gamma_0$  exhibit a similar flat `plateau', as previously discussed.

Next, we estimate the average time for particles to reach $\gamma_{f} > 3\gamma_0$. We observe that the average energy increases through time as $\propto \sqrt{t}$ after hitting the shock. This scaling is consistent with particle diffusion in the small-angle scattering regime demonstrated in relativistic pair shocks \citep{Sironi_2013, Plotnikov_2018}. In this regime, the average energy of the injected particles grows as $\propto \sqrt{t}$. The shell hits the shock at $\omega_{p}t \sim 1276$, and in the small-angle scattering regime, the average energy of the particles moving along with the shock should reach $\gamma \sim 3.2\gamma_0$ at $\omega_{p}t \sim 1465$. This prediction is consistent with the true average energy gain through time of our tracked shell. We also note that this time interval is consistent with the duration of the signature of the pre-acceleration phase (i.e., the flat `plateau' shared by curves with $\gamma_f > 3\gamma_0$) in Figure \ref{percentage_plot}.

This scaling thus yields an approximate time required to reach the free-streaming stage of $\omega_p \Delta T \sim 189$. We now estimate the percentage of particles left in the shock transition region at $\omega_p \Delta T \sim 189$.

The fraction of escaping particles depends on two main parameters. First, the rate of filament merging or kink at the shock transition that defines the rate at which reflected particles must switch from one filament to another. Second, the probability of survival of a given particle as it switches from one filament to another,  i.e., the likelihood of a particle to find another filament and continue heading in  the $+x$ direction. Since filament switches are rare and independent, we approximate the number $n$ of switches between filaments as a Poisson distribution, with probability mass function ${P(n) = \lambda^n e^{-\lambda}/n!}$, where $\lambda$ is the average rate of
filament switching for a given interval of time. From the high-energy particle orbits in the time interval $\omega_{p}t \sim 1276 - 1465$, we obtain that most high-energy particles had to switch filaments four distinct times. We thus estimate a rate of $\lambda = 4$ filament switches per $\omega_p \Delta  T$, where $\omega_p \Delta  T = 189$.

Then, we denote the probability that a particle will find another filament when a switch occurs as $p$. In the following calculation, we use $p = 0.5$, since by symmetry, at any given moment, the shock upstream is composed equally of regions of positive and negative current.

Now, we estimate the fraction of reflected particles moving along with the shock after $\omega_p \Delta T$ by performing a weighted average over the different possible occurrences of filament switches that a particle may encounter:
\begin{equation}
\begin{split}
    f_{\mathrm{survival}} = \sum_{n=0}^{\infty} P(n)p^n = e^{\lambda(p-1)} \approx 0.14
\end{split}
\end{equation}

Thus, we obtain that $\sim$14\% of reflected particles survive a sequence of filament switches over a time $\omega_p \Delta T$. We approximated earlier that $\sim$35\% of incoming particles are reflected in the first place, therefore we estimate that $\sim$5\% of the total incoming particles remain with the shock long enough to reach the energy required to enter the free-streaming phase. Comparing this quantity to the $\sim$1\% of particles in the high-energy tail of the particle distribution, we find reasonable agreement, given our approximations.

It is worth noting that we do not expect perfect agreement since the two percentages we are comparing here are subtly different. Notably, the $\sim$1\% comes from the nonthermal tail of the \emph{downstream} particle distribution. By the end of the simulation, most of the $\sim$5\% of particles described above are still participating in DSA and have not yet entered the downstream. This being said, our estimate of $\sim$5\% captures the essential physics at play that determine the $\sim$1\%.

\section{Discussion and Conclusions} \label{sec:disc_conclu}

To summarize, Weibel-mediated shocks studied through PIC simulations accelerate $\sim$1\% of incoming particles, while the majority simply pass through the shock and become thermalized. In order to determine the physics at play behind this fraction of high-energy particles, we studied the microphysics of particle reflection and transmission in relativistic pair shocks. We found that particle reflection in Weibel-mediated shocks is not solely a random, turbulent process, but rather that there exist local structures on the shock which are efficient at transmitting one species of particles and reflecting the other. We developed a toy model for these reflective structures produced by the approach of filaments of opposite currents, which allowed us to understand the observation that every positron in an electron filament and every electron in a positron filament hitting the shock is reflected, while the rest of the particles are transmitted. This insight then allowed us to estimate the percentage of particles which are initially reflected at the shock, which we found to be $\sim$35\%. 
The subsequent survival of reflected particles depends on their ability to find current filaments of the right sign where they can get guided away from the shock without being swept downstream. This non-diffusive transport is reset every time the filament ends, and the particle needs to jump to another properly signed filament. This pre-acceleration continues until the particle gains enough energy from scatterings to be able to diffuse through the upstream and join the DSA.  The filament jumping process is lossy, however, and we constructed a probabilistic survival model which allowed us to estimate the percentage of remaining high-energy particles at  $\sim$5\%. The multiplicative two-step probabilistic process (shock reflection probability and filament survival probability) 
captures the essential microphysics that gives rise to the $\sim$1\% of particles in the high-energy tail of the downstream particle distribution observed in simulations.

The discrepancy between the percentage of high-energy particles we estimate based on our model ($\sim$5\%) and the percentage of particles observed in the high-energy tail of the downstream particle distribution ($\sim$1\%) is likely due to our estimates of $\lambda$, the average rate of filament switches, and $p$, the probability of survival of a particle when encountering a switch in filament. These rates and probabilities were estimated by inspection of a large sample of test particles, as it is nontrivial to algorithmically define the criteria for finding filament trapping and scattering. Indeed, we note that if $p$ were slightly smaller ($p \sim 0.4$) and $\lambda$ were slightly larger ($\lambda \sim 5$ filament switches per $\omega_p\Delta T$), the resulting percentage of high-energy particles would be $\sim$1.7\%. This being said, a more refined measurement of these two quantities requires a clearer picture of non-diffusive particle transport in filamentary turbulence, which we leave to future work. 

Our model also allows us to predict the impact of changes in the properties of Weibel filamentation on the acceleration efficiency. For example, the fraction of high-energy particles depends on the charge separation of the filaments as they hit the shock. Indeed, we predict that the acceleration efficiency should drop with time as the filaments become more and more charge separated, since a smaller percentage of particles would be reflected at the shock during the first crossing. This process could provide a feedback mechanism that regulates the number of accelerated particles and their effect on the flow at later times. 

The long-term evolution of Weibel-mediated shocks remains underexplored. Indeed, it is possible that changes in the properties of the upstream turbulence may have greater effects on the acceleration efficiency than the gradual increase in charge-separation. For example, at later times, electromagnetic structures in the upstream may merge to form larger and larger structures \citep{Keshet_2009}, thus creating significant density voids near the shock. The creation of density voids of a larger scale than the background filaments would help beam particles survive the pre-acceleration phase. This is because reflected particles need to find regions of the same sign of current in order to stay with the shock, regardless of whether the current region is associated with a density filament or not. Since these voids may be on a larger scale than the filaments, we expect them to be populated with beam particles and to break up less frequently than the smaller-scale density filaments. Beam particles would thus have to switch current regions fewer times, leading to a lower average rate of switching $\lambda$ and a higher percentage of beam particles which survive the pre-acceleration phase.

In addition, the assumption that the shock upstream is composed equally of regions of positive and negative currents may not hold for the long-term evolution of the shock. If an asymmetry in the upstream current regions arises, then the probability of survival $p$ will become different for each beam species, favoring the acceleration of one beam species over the other. Particles of the favored beam species will sit on the leading edge of the shock in these over-represented regions of the same sign of current, thus further enhancing the current. However, this situation cannot continue forever, as background particles of the same species as the favored beam species will be expelled from the current region, leading to an increase in charge separation, and fewer initially reflected particles. We thus expect that an asymmetry in the upstream current regions would lead initially to an increase in acceleration efficiency for one species and a decrease for the other species, before self-regulating by reducing the percentage of initially reflected particles of the favored species. This could lead to long term oscillatory behavior in the shock structure. 

Further study of the pre-acceleration phase may be enlightening. One interesting possibility would be to model the pre-acceleration phase as a ``first-passage process," in order to obtain a fully analytic description for the probability that a returning particle first reaches a certain energy threshold at a specific time (see, e.g., \citealt{Redner_2022}). In the relativistic regime of electron-ion Weibel-mediated shocks, energy equipartition between species at the shock transition~\citep{Spitkovsky_2008, Haugbolle_2011, Gedalin_2012, Plotnikov_2013, Kumar_2015, Vanthieghem_2022} significantly reduces the difference in inertia between electrons and ions. Consequently, deviations from the current model are anticipated to be minor, resulting in comparable injection rates for both electrons and ions. Such a description is left for further studies.

\smallskip

We thank T. Dyson for helpful comments on the manuscript, and E. Nakar and M. Medvedev for useful discussions. We acknowledge  support from by the Multimessenger Plasma Physics Center (MPPC, NSF grant
PHY-2206607), NASA grant 80NSSC20K1273, the Simons Foundation (grant 267233), and the NIFS Collaboration Research Program (NIFS22KIST020).

\vspace{5mm}

\pagebreak
\bibliography{sample631}{}
\bibliographystyle{aasjournal}

\end{document}